\documentclass[secnumarabic, graphics,floatfix, nofootinbib,tightenlines,nobibnotes, aps, prl, 12pt]{revtex4-2}
\usepackage[english]{babel}
\usepackage[utf8]{inputenc}
\usepackage{amsfonts}
\usepackage{multirow}
\usepackage{graphicx}
\usepackage{epstopdf}
\usepackage{hyperref}
\usepackage{latexsym}
\usepackage{amsmath}
\usepackage{amssymb}

\usepackage[utf8]{inputenc}
\hypersetup{
    colorlinks=true,
    linkcolor=red,
    citecolor=blue,
}
\usepackage{color}
\usepackage[T1]{fontenc}
\usepackage{txfonts}
\usepackage{mathrsfs}
\usepackage[center]{subfigure}

\begin{document}
\setcounter{secnumdepth}{2}
\renewcommand{\theequation}{1.\arabic{equation}} \setcounter{equation}{0}

\newcommand{\bq}{\begin{equation}}
\newcommand{\eq}{\end{equation}}
\newcommand{\bqn}{\begin{eqnarray}}
\newcommand{\eqn}{\end{eqnarray}}
\newcommand{\nb}{\nonumber}
\newcommand{\lb}{\label}
\newcommand{\be}{\begin{equation}}
\newcommand{\en}{\end{equation}}
\newcommand{\PRL}{Phys. Rev. Lett.}
\newcommand{\PL}{Phys. Lett.}
\newcommand{\PR}{Phys. Rev.}
\newcommand{\CQG}{Class. Quantum Grav.}
 
\title{Standing wave in perturbed anti-de Sitter spacetimes with a naked singularity}

\author{Kai Lin$^{1,2}$}\email[E-mail: ]{lk314159@hotmail.com}
\author{Wei-Liang Qian$^{2,3,4}$}\email[E-mail: ]{wlqian@usp.br}

\affiliation{$^{1}$ Hubei Subsurface Multi-scale Imaging Key Laboratory, School of Geophysics and Geomatics, China University of Geosciences, 430074, Wuhan, Hubei, China}
\affiliation{$^{2}$ Escola de Engenharia de Lorena, Universidade de S\~ao Paulo, 12602-810, Lorena, SP, Brazil}
\affiliation{$^{3}$ Faculdade de Engenharia de Guaratinguet\'a, Universidade Estadual Paulista, 12516-410, Guaratinguet\'a, SP, Brazil}
\affiliation{$^{4}$ Center for Gravitation and Cosmology, School of Physical Science and Technology, Yangzhou University, 225002, Yangzhou, Jiangsu, China}

\date{May. 9th, 2024}

\begin{abstract}
    In the framework of black hole perturbation theory, this work investigates the standing wave solutions in Reissner-Nordtsr\"om (RN) anti-de Sitter (AdS) spacetimes with a naked singularity. 
    These solutions can be viewed as a specific class of quasinormal modes exhibiting distinct characteristics.
    The imaginary parts of their frequencies are numerically vanishing, allowing them to persist over an extended period.
    Besides, these modes are predominantly stationary in terms of the evolution of spacetime waveforms.
    The numerical calculations are carried out employing the finite difference method, and the quasinormal frequencies extracted by the Prony method are shown to be consistent with those obtained using the matrix method.
    The obtained waveforms and quasinormal frequencies are shown to be drastically different from those of an extreme RN-AdS black hole.
    As the quasinormal modes are primarily dissipative, the non-dissipative standing waves are attributed to the nature that the singularity can neither be a sink nor a source of the gravitational system.
\end{abstract}


\maketitle
\newpage

\section{Introduction}\label{section1}

The revolutionary advent of gravitational wave (GW) detection, made possible by the direct measurement of the waveforms~\cite{agr-LIGO-01, agr-LIGO-02, agr-LIGO-03, agr-LIGO-04}, has heralded a new era of observational astronomy. 
This pioneering field of research seeks to probe the strong-field regime of gravity, a pursuit that has seen rapid advancements in recent years. 
In particular, the ongoing space-borne laser interferometer projects~\cite{agr-LISA-01, agr-TianQin-01, agr-Taiji-01} have promoted much efforts aiming at unprecidented detector sensitivity~\cite{agr-TDI-review-01, agr-TDI-review-02, agr-TDI-Wang-03, agr-SNR-Wang-01, agr-SNR-Wang-02}.

Theoretical perspectives have led to essential estimations on the viability of black hole spectroscopy~\cite{agr-bh-spectroscopy-05, agr-bh-spectroscopy-06, agr-bh-spectroscopy-15, agr-bh-spectroscopy-18, agr-bh-spectroscopy-20, agr-bh-spectroscopy-36}. 
In particular, for real-world scenarios, the gravitational radiation sources, such as black holes or neutron stars, are not isolated but interact with surrounding matter. 
In other words, there is inevitably a deviation of spacetime from an ideally symmetric metric, causing the emitted GWs to differ substantially from those of a pristine, isolated, compact object. 
This phenomenon has steered investigations towards the study of ``dirty'' black holes~\cite{agr-bh-thermodynamics-12, agr-qnm-33, agr-qnm-34, agr-qnm-54}, opening new avenues in black hole perturbation theory.

In this regard, researchers have focused on modeling realistic systems composed of compact astrophysical objects, such as black holes or neutron star binaries. 
A key focus has been the study of black hole quasinormal modes (QNMs)~\cite{agr-qnm-review-02, agr-qnm-review-03, agr-qnm-review-06}, which are integral to the ringdown stage of the merger process. 
These dissipative oscillations, embodying the inherent properties of the underlying black hole spacetime, are governed by several no-hair theorems~\cite{agr-bh-nohair-01, agr-bh-nohair-04}.
Leung {\it et al.} pioneered the study of scalar QNMs of dirty black holes for nonrotating metrics, assessing the deviations in quasinormal frequencies using the generalized logarithmic perturbation theory. 
Subsequently, Barausse {\it et al.} performed a comprehensive analysis concerning perturbations around a central Schwarzschild black hole~\cite{agr-qnm-54}. 
Their observations confirmed that the resultant QNMs could significantly deviate from those of an isolated black hole. 
However, they also concluded that the astrophysical environment would not significantly affect black hole spectroscopy if an appropriate waveform template was utilized. 
Of the various scenarios explored in~\cite{agr-qnm-54}, the thin shell model emerged as the one providing the most substantial modification to the QNM spectrum.
Meanwhile, the backreaction to the metric may also play a pertinent role~\cite{Babichev:2012sg, Nakamura:2021mfv, Kimura:2021dsa, deCesare:2022aoe, deCesare:2023rmg}. 
The concept of black hole pseudospectrum, pioneered by Nollert and Price~\cite{agr-qnm-35, agr-qnm-36}, is closely connected with the considerations discussed above. 
They demonstrated that minor perturbations, expressed as step functions, significantly impact high-overtone modes in the QNM spectrum. 
This demonstrated an unexpected instability of the QNM spectrum against ``ultraviolet'' perturbations, challenging the assumption that a reasonable approximation of the effective potential ensures minimal deviation in the resulting QNMs. 
In~\cite{agr-qnm-lq-03}, we argued that even in the presence of discontinuity, the asymptotic behavior of the QNM spectrum would be non-perturbatively modified. 
Specifically, high-overtone modes would shift along the real axis instead of ascending the imaginary frequency axis~\cite{agr-qnm-continued-fraction-02, agr-qnm-continued-fraction-03}. 
We found that this phenomenon persisted regardless of the discontinuity's distance from the horizon or its significance.
Employing the concept of structural stability, Jaramillo {\it et al.}~\cite{agr-qnm-instability-07, agr-qnm-instability-13, agr-qnm-instability-14} analyzed the problem in the context of randomized perturbations to the metric. 
Their analysis, in conjunction with Chebyshev's spectral method in hyperboloidal coordinates~\cite{agr-qnm-hyperboloidal-01}, revealed that the boundary of the pseudospectrum moves toward the real frequency axis. 
These results reinforce the universal instability of high-overtone modes triggered by ultraviolet perturbations.
More recent observations by Cheung {\it et al.}~\cite{agr-qnm-instability-15} indicate that even the fundamental mode can be destabilized under generic perturbations.

A naked singularity poses a challenge due to its impact on the causal structure of the underlying spacetime manifold.
Specifically, due to its presence, the spacetime in question mostly does not contain a complete Cauchy surface.
From a physical perspective, it is intriguing that the requirement for geodesic completeness~\cite{book-cosmology-Hawking} can be replaced by global hyperbolicity~\cite{book-general-relativity-singularity-Clarke, agr-singularity-60}.
The difficulty caused by singularity is somewhat mitigated by shifting the focus from the classical point particles to the fields~\cite{agr-singularity-25, agr-singularity-26, agr-singularity-27, agr-singularity-28, agr-singularity-29, agr-singularity-40, agr-singularity-50}, as the wave might propagate through the would-be singularity in a well-defined fashion.
Nonetheless, in a non-globally hyperbolic spacetime, the uniqueness of solutions of a wave equation is lost.
As a result, an initial value problem is ill-defined when the time evolution of the field is not uniquely determined in the entire spacetime domain.
A possible solution is introducing additional boundary condition(s) so that the wave propagation is uniquely defined~\cite{agr-singularity-26, agr-singularity-28, agr-singularity-40, agr-singularity-52}.
Alternatively, in terms of spectral theorem~\cite{book-methods-mathematical-physics-08}, the problem can be viewed as the self-adjointness of the positive and symmetric Hamiltonian operator comprised of spatial operation in the master equation. 
To be specific, by suitably prescribing its domain~\cite{agr-singularity-26, agr-singularity-28}, such as Friedrich's extension~\cite{book-methods-mathematical-physics-08}, the operator in question can be shown to be positive symmetric self-adjoint under moderate assumptions.
Subsequently, the self-adjoint extension acts as a time translation operator that uniquely determines the system's time evolution.
Effectively, any undesirable singular mode not contained in the domain of the initial data will not appear after the scattering process, so the arbitrariness in time evolution is removed.
Subsequently, by specifying the dynamics using such a recipe, one can characterize the system by a set of quasinormal modes.

The present study is motivated to explore the properties of the QNMs in spacetime with a naked singularity.
We report the occurrence of a specific class of quasinormal frequencies, which demonstrate themselves as standing wave solutions in the Reissner-Nordtsr\"om (RN) anti-de Sitter (AdS) spacetimes.
Regarding the evolution of spacetime waveforms, these modes are primarily stationary. 
Their characteristics are distinct from the QNMs of extreme RN-AdS black holes, even though the underlying spacetime metrics are closely related.
Specifically, the causal one-way membrane turns into an antinode of the resulting standing wave.
We explore the spatial-temporal evolution of the perturbations and derive the specific values of the corresponding frequencies.

The remainder of the paper is organized as follows.
The following section gives an account of the master equation of the axial oscillations in the RN-AdS spacetime with a naked singularity.
In Sec.~\ref{section3}, the spatial-temporal evolutions are evaluated numerically using the finite difference method.
The corresponding quasinormal frequencies are extracted using the Prony method and compared to those obtained using the matrix method.
By comparing the QNMs of the corresponding extreme black holes, in Sec.~\ref{section4}, we elaborate on the difference between the two scenarios.
Further discussions and the concluding remarks are given in Sec.~\ref{section5}.

\section{The master equation in the Reissner-Nordtsr\"om anti-de Sitter spacetime with a naked singularity}\label{section2}
\renewcommand{\theequation}{2.\arabic{equation}} \setcounter{equation}{0}

By solving the Einstein field equation in the presence of a cosmological constant $\Lambda$
\bq \label{EFEquation}
G_{\mu\nu}+\Lambda g_{\mu\nu} =0 ,
\eq
the metric of a charged (anti-)de Sitter black hole is found to be~\cite{agr-bh-Kerr-10}
\bq
\label{Metric1}
ds^2=-f(r)dt^2+\frac{dr^2}{f(r)}+r^2\left(d\theta^2+\sin^2\theta d\varphi^2\right) ,
\eq
where
\bq
\label{Metric2}
f(r)=1-\frac{2M}{r}+\frac{Q^2}{r^2}-\frac{\Lambda}{3}r^2 ,
\eq
where $M$ and $Q$ are the mass and electric charge of the solution.

For the present study, we are interested in the AdS spacetime $\Lambda<0$.
It is readily verified when 
\bqn
M<\frac{\sqrt{2-2\sqrt{1-4Q^2\Lambda}}}{6\sqrt{\Lambda}}\left(2+\sqrt{1-4Q^2\Lambda}\right) ,\label{condSing}
\eqn
the metric function $f(r)$ does not have any positive root and, subsequently, $r=0$ becomes a naked singularity.

The master equation for perturbations in the above spacetime is found to have the form
\bq
\label{QNMs1}
\frac{\partial^2\Psi}{\partial r_*^2}-\frac{\partial^2\Psi}{\partial t^2}-V(r)\Psi=0
\eq
where $r_*=\int{dr/f}$ is the tortoise coordinate.
If one choose $r_*^0\equiv r_*(r=0)=0$, then $r_*^\infty\equiv r_*(r\rightarrow\infty)$ is manifestly finite. 
The effective potential $V(r)$ in Eq.~\eqref{QNMs1} for the spherically symmetric axial gravitational and Maxwell perturbations are given by~\cite{book-blackhole-Chandrasekhar}
\bqn
\label{QNMs2}
V_\text{axial}^\pm&=&\frac{f(r)}{r^3}\left[2(n_a+1)r-p_{a}^\mp\left(1+\frac{p_a^\pm}{2n_ar}\right)\right] ,
\eqn
where 
\bqn
p_a^\pm&=&3M\pm\sqrt{9M^2+8n_aQ^2},\nb\\
n_a&=&\frac{(L-1)(L+2)}{2} .
\eqn
It is noted that in an RN spacetime, the perturbed electromagnetic field is related to the perturbed components of the Ricci tensor.
In the spherically symmetric axial case, there are only two independent degrees of freedom, and they can be decoupled, giving rise to the effective potential given by Eq.~\eqref{QNMs2}.
Without loss of generosity, we will use the parameters $\Lambda=-3$ and $L=3$ in the remainder of the paper.
For the case of naked singularity, we mainly adopt $M=3$ and $Q=\sqrt{5}$.

\begin{figure*}
\includegraphics[width=8cm]{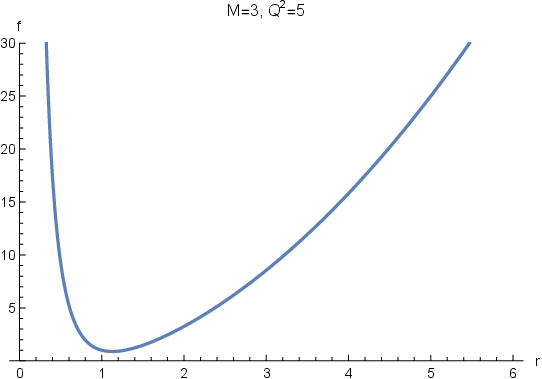}\includegraphics[width=8cm]{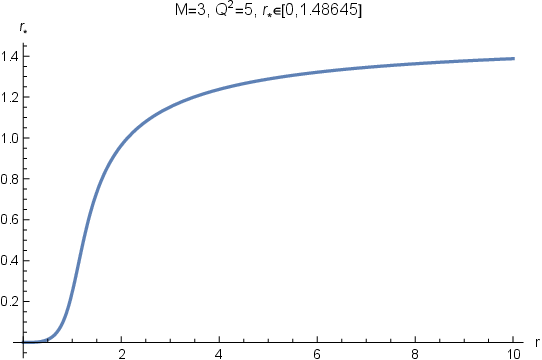}
\includegraphics[width=8cm]{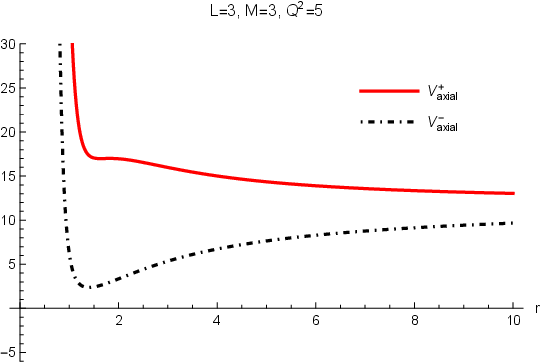}\includegraphics[width=8cm]{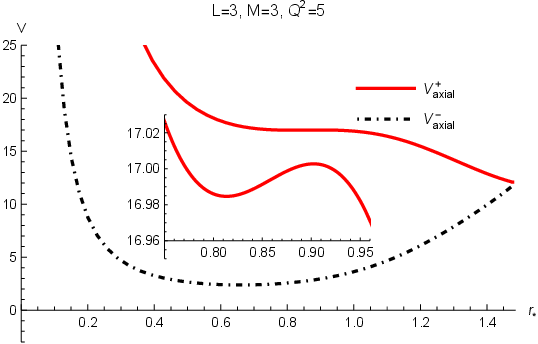}
\caption{The metric function $f(r)$ (top-left), tortoise coordinate $r_*(r)$ (top-right) and the effective potentials $V^\pm_\mathrm{axial}$ for the RN-AdS spacetime with a naked singularity shown in radial (bottom-left) and tortoise coordinates (bottom-right).
The effective potential $V^-_\mathrm{axial}$ features a local minimum that is less than the potential at spatial infinity.
In the calculations, we have used the parameters $M=3$, $Q=\sqrt{5}$, $\Lambda=-3$, and $L=3$, which is primarily adopted by the remainder of the paper in the case of a naked singularity.} \label{fig1}
\end{figure*}

In Fig.~\ref{fig1}, we illustrate the effective potentials $V(r)$ for the axial perturbations and the metric function $f(r)$.
Observing Eqs.~\eqref{QNMs2}, both effective potentials possess the limits 
\bqn
\lim\limits_{x\to 0}V_\text{axial}^\pm (x) &=& +\infty \nb\\
\lim\limits_{x\to +\infty}V_\text{axial}^\pm (x) &=&  L(L+1) .
\eqn
Also, it is noted that the effective potential $V = V^-_\mathrm{axial}$ features a local minimum forming a potential well, whose value is less than that at spatial infinity.
This allows for standing wave solutions elaborated below.
In what follows, we will primarily focus on the axial degree of freedom $\Psi=\Psi^-_\mathrm{axial}$ related to the effective potential $V = V^-_\mathrm{axial}$.

Owing to the singularity at $r=0$, the initial value problem is ill-defined as the spacetime is not globally hyperbolic.
To proceed, we follow the recipe and arguments elaborated in~\cite{agr-singularity-52} by choosing a physically relevant boundary condition at the singularity.
Subsequently, Green's function can be constructed using the solutions of the corresponding homogeneous equation~\cite{agr-qnm-review-02}.
Therefore, the system's resulting QNMs are governed by the poles of the Green's function, which, in turn, can be obtained using most standard approaches.
To this end, we first analyze the asymptotic behavior of the waveforms in the vicinity of the origin.
\bqn
\label{QNMs3b}
\Psi(r\rightarrow 0)&\rightarrow & C_1r^4+\frac{C_2}{r},
\eqn
where $C_1$ and $C_2$ are two constants of integration.
It is a natural choice to demand that the flow at the origin vanishes, namely,
\bqn
\label{QNMJ0}
j_r(r\rightarrow 0)= \left.\left[\Psi(r,t)\partial_r\Psi^*(r,t)-\Psi^*(r,t)\partial_r\Psi(r,t)\right]\right|_{r\to 0}=0,
\eqn
This demands that the singularity is neither a source nor a sink for the field, and subsequently, the $S$ matrix of the field is a unitary operator.
Furthermore, we adopt the boundary condition $C_2=0$, namely,
\bqn
\label{bc_ho}
\Psi(0,t)=0 .
\eqn
One may argue that it is a relevant choice.
On the one hand, it implies Eq.~\eqref{QNMJ0}.
On the other hand, it can be shown~\cite{agr-singularity-26, agr-singularity-52} that such a boundary condition corresponds to Friedrich's extension of a symmetric operator when singularity is produced by ``cutting out holes''.
We note that such a choice of boundary condition is not unique and can be replaced by another physically meaningful recipe.
For the standing wave solution, the boundary condition at infinity is taken to be
\bqn
\label{bc_inf}
\Psi(+\infty,t)=0 .
\eqn

Now, Green's function can be written down by using the two solutions of the homogeneous equation that satisfy the boundary conditions at $r=0$ and $+\infty$~\cite{agr-qnm-review-02}, namely,
\begin{equation}
\label{FormalGreen}
\widetilde{G}(x,y;\omega)= \frac{1}{W(\omega)}\Psi_0(x_<,\omega)\Psi_\mathrm{inf}(x_>,\omega) ,
\end{equation}
where $x_<\equiv \min(x, y)$, $x_>\equiv \max(x, y)$, and
\begin{equation}
\label{DefWronskian}
W(\omega) \equiv W(\Psi_\mathrm{inf}, \Psi_0) = {\Psi_\mathrm{inf}} {\Psi_0}' - {\Psi_0} {\Psi_\mathrm{inf}}' 
\end{equation}
is the Wronskian, where $\Psi_0$ and $\Psi_\mathrm{inf}$ are the two linearly independent solutions of the corresponding homogeneous equation satisfying the boundary conditions Eqs.~\eqref{bc_ho} and~\eqref{bc_inf} at the horizon and infinity.
Therefore, the above procedure uniquely defines Green's function, whose poles give rise to the quasinormal modes.
In what follows, Eq.~\eqref{QNMs1} and boundary conditions Eqs.~\eqref{bc_ho} and~\eqref{bc_inf} will be employed to calculate the spatial-temporal evolutions of the perturbations using the finite difference method.
The corresponding quasinormal frequencies will also be evaluated using the matrix method.
For the latter, we introduce the transformation $\Psi=e^{-i\omega t}\Psi(r)$, where $\omega$ is the waveform frequency.

\section{Spatial-temporal evolutions of the perturbations and the standing wave solutions}\label{section3}

In this section, we investigate the spatial-temporal evolution of the gravitational perturbations and explore the underlying quasinormal frequencies.
We employ the finite difference method to evaluate the spatial-temporal dependence of the waveforms, and the matrix method is utilized to study both the waveforms and the resonance frequencies more specifically.
Also, the Prony method is used to extract the frequencies from the temporal profiles, and the obtained results are compared with those obtained by the matrix method.
We further elaborate on a possible transition between two non-dissipative fundamental modes in the extreme RN-AdS black hole metric and that with a naked singularity, whose metric forms are essentially identical outside the ``horizon''.

The finite difference method~\cite{agr-qnm-finite-difference-01, agr-qnm-finite-difference-02, agr-qnm-finite-difference-03, agr-qnm-finite-difference-04, agr-qnm-finite-difference-05} serves as an effective tool for studying the complex field of gravitational waveforms.
This method offers a more general description of the dynamical evolution for given initial perturbations. 
The inherent flexibility of the finite difference method allows it to accommodate diverse scenarios and boundary conditions, making it particularly useful in studying the propagation and interaction of gravitational waves.

\begin{figure}
\includegraphics[width=8cm]{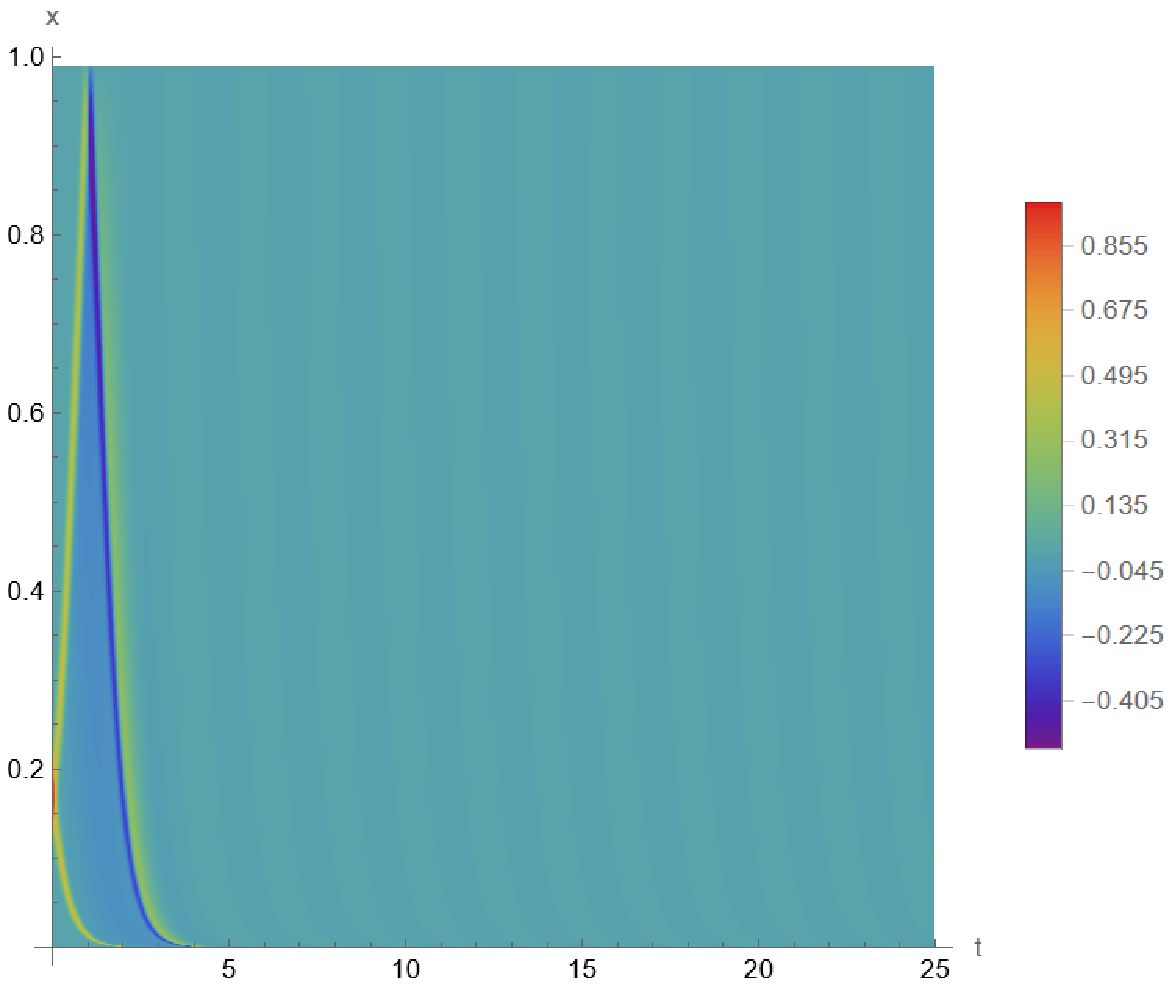}\includegraphics[width=8cm]{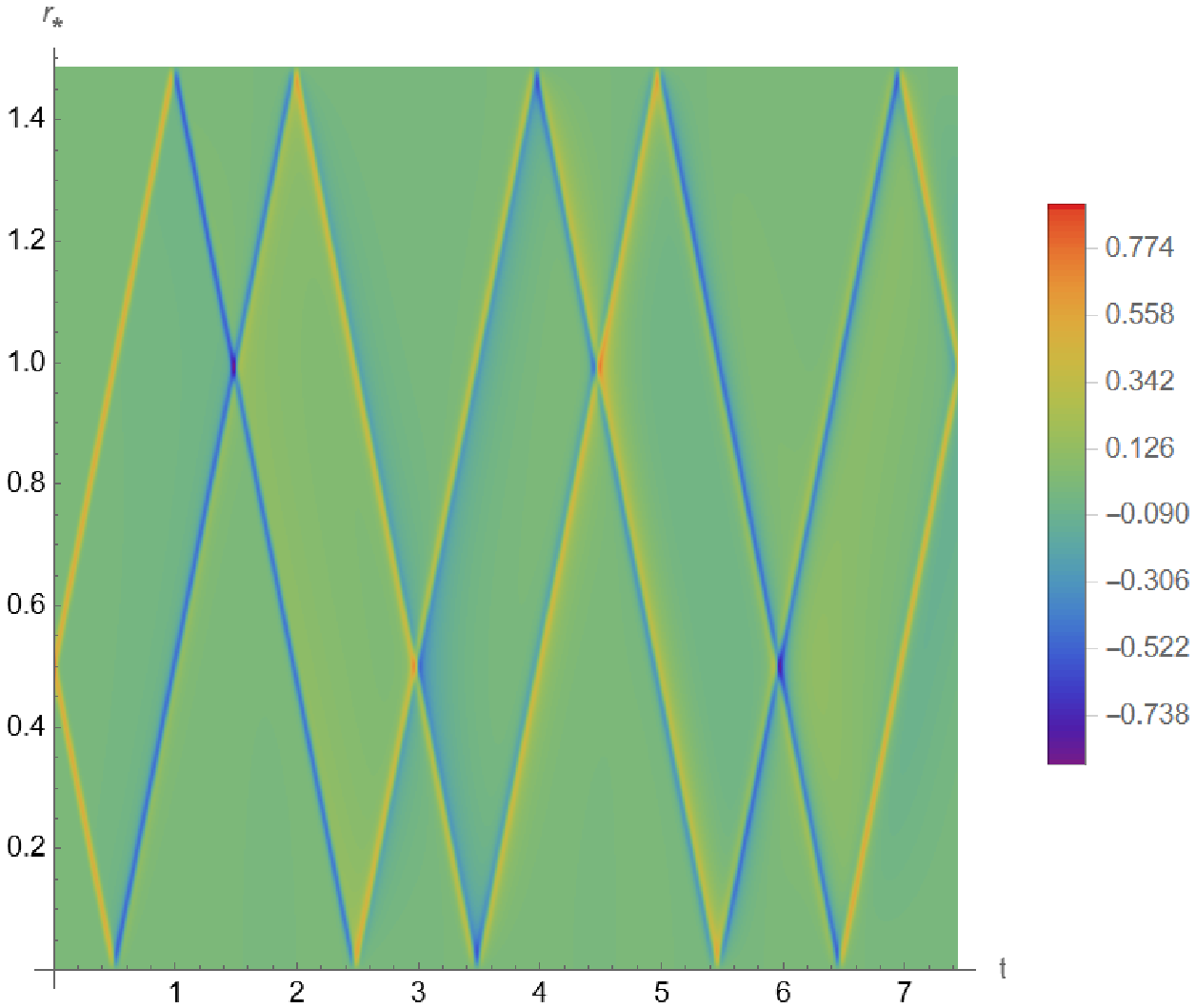}
\caption{The space-temporal evolution of the initial axial perturbations of a Gaussian form in an RN-AdS black hole metric (left) and that with a naked singularity (right).} \label{fig2}
\end{figure}

\begin{figure}
\includegraphics[width=8cm]{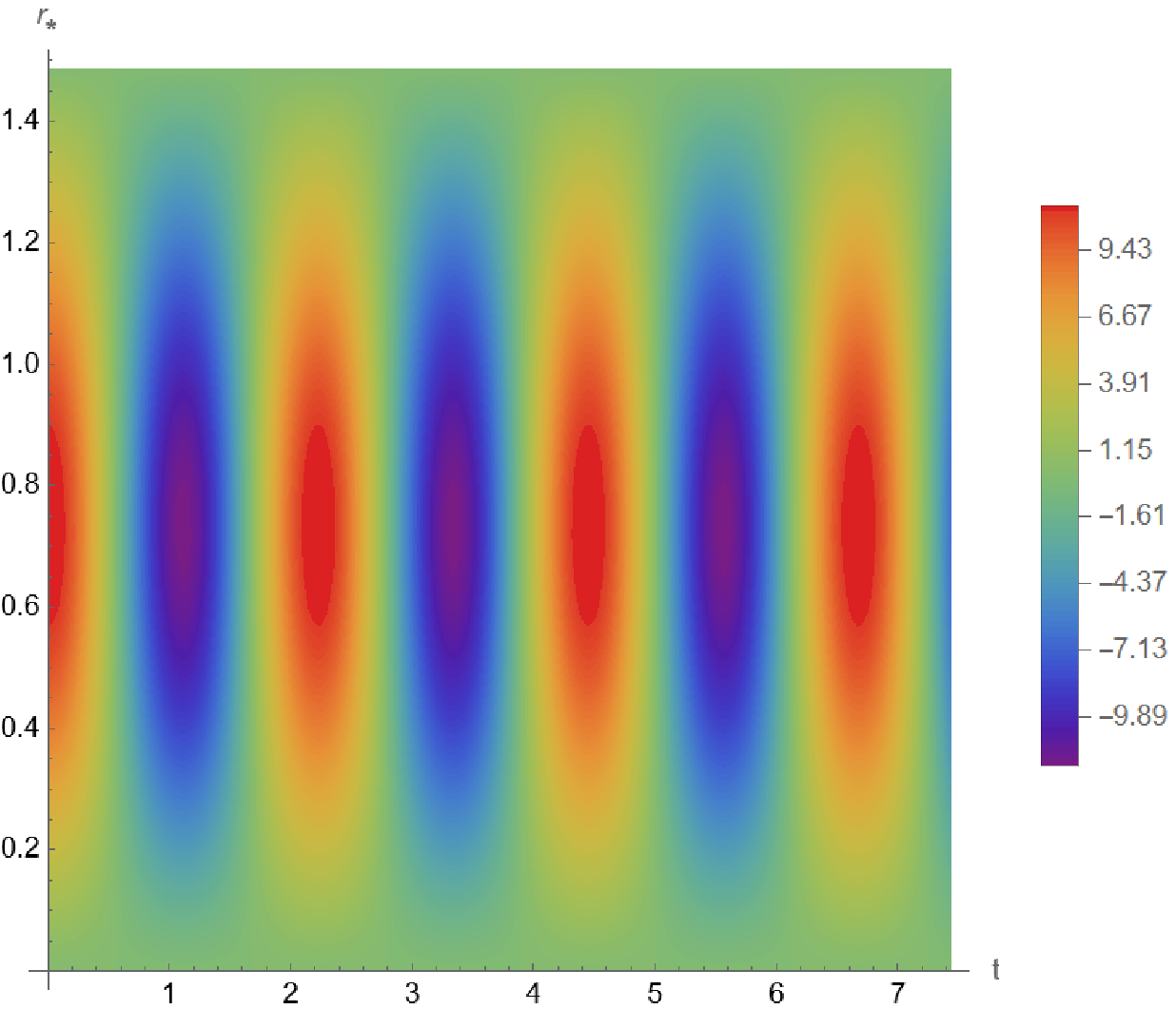}\includegraphics[width=8cm]{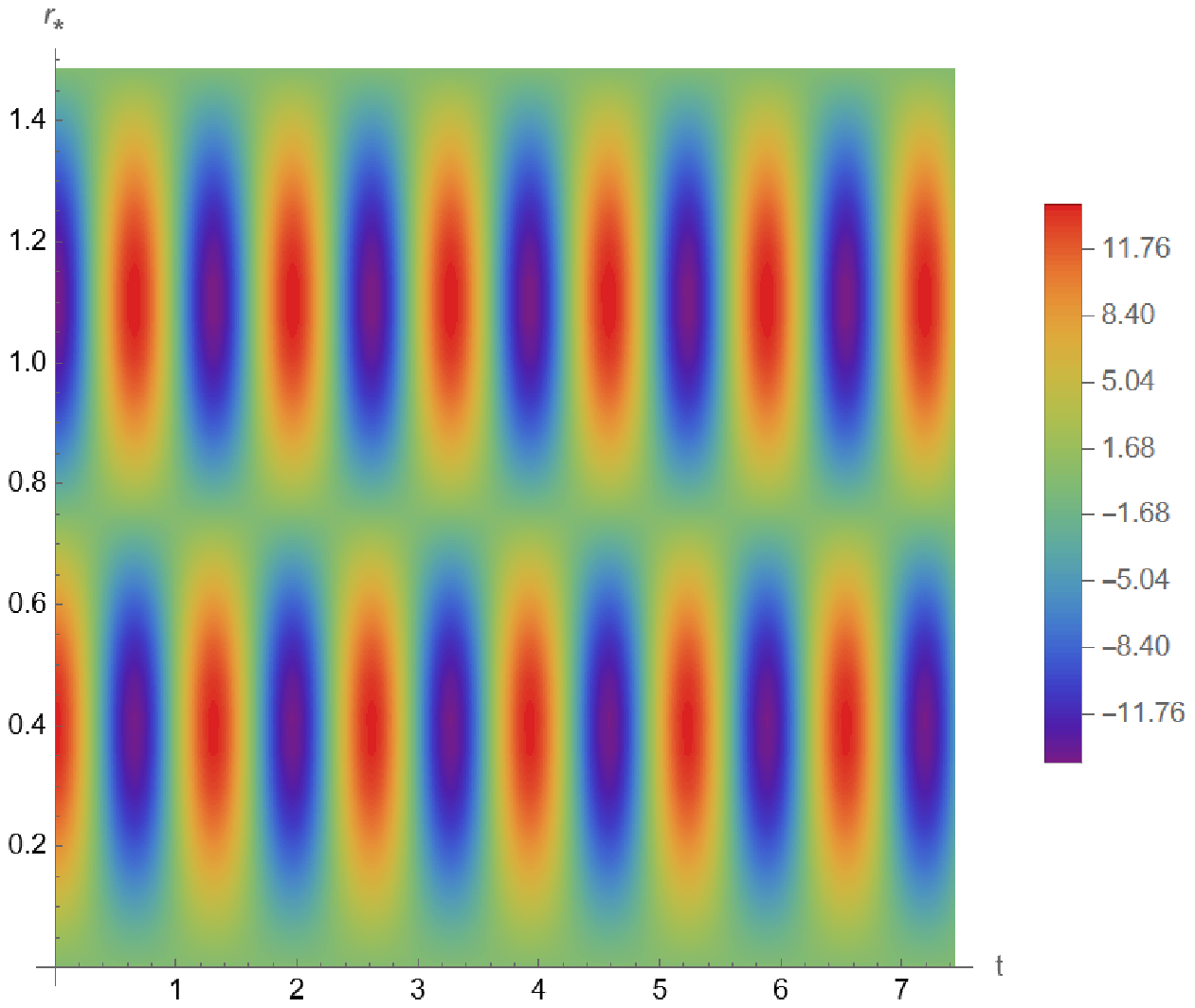}
\includegraphics[width=8cm]{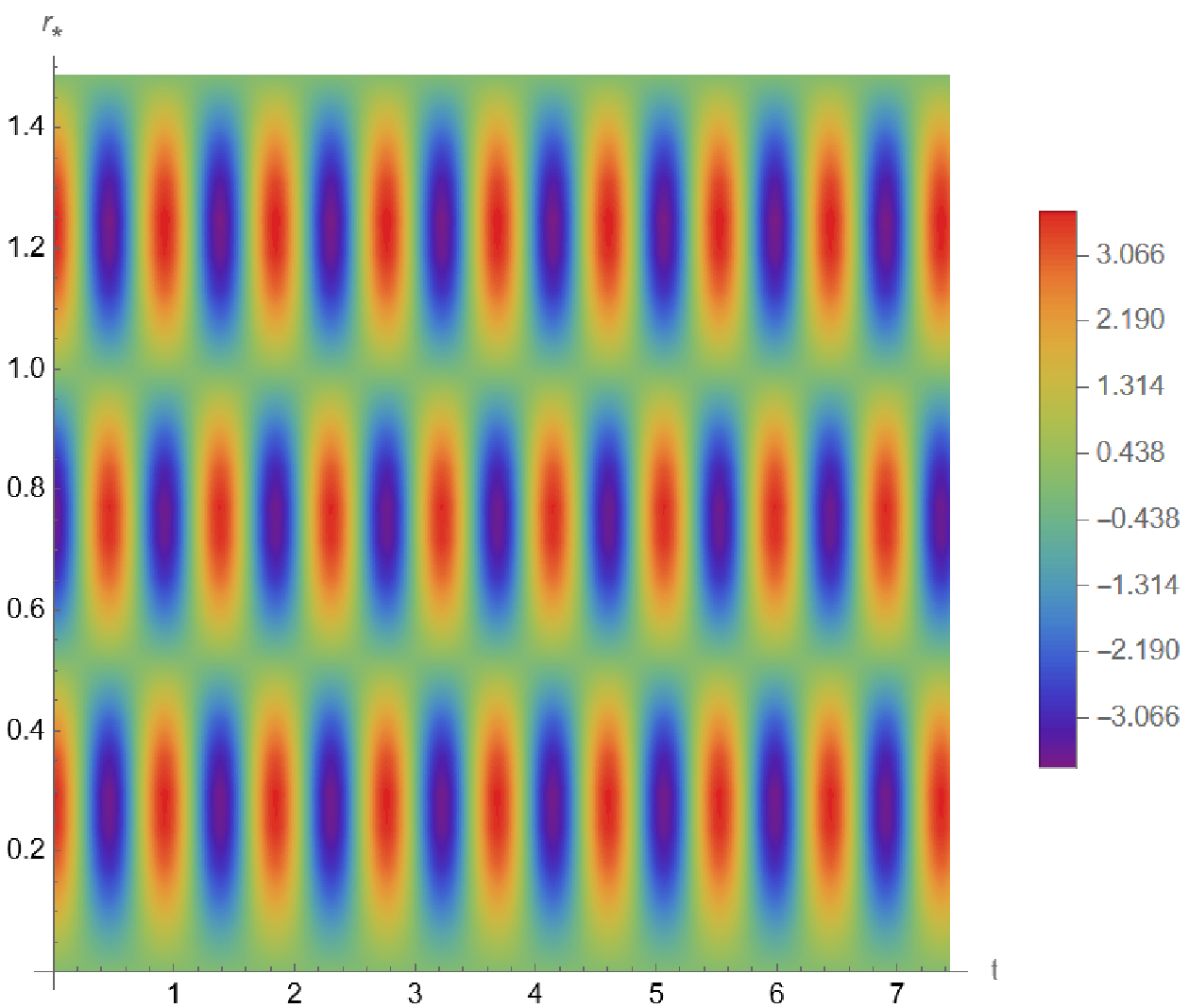}\includegraphics[width=8cm]{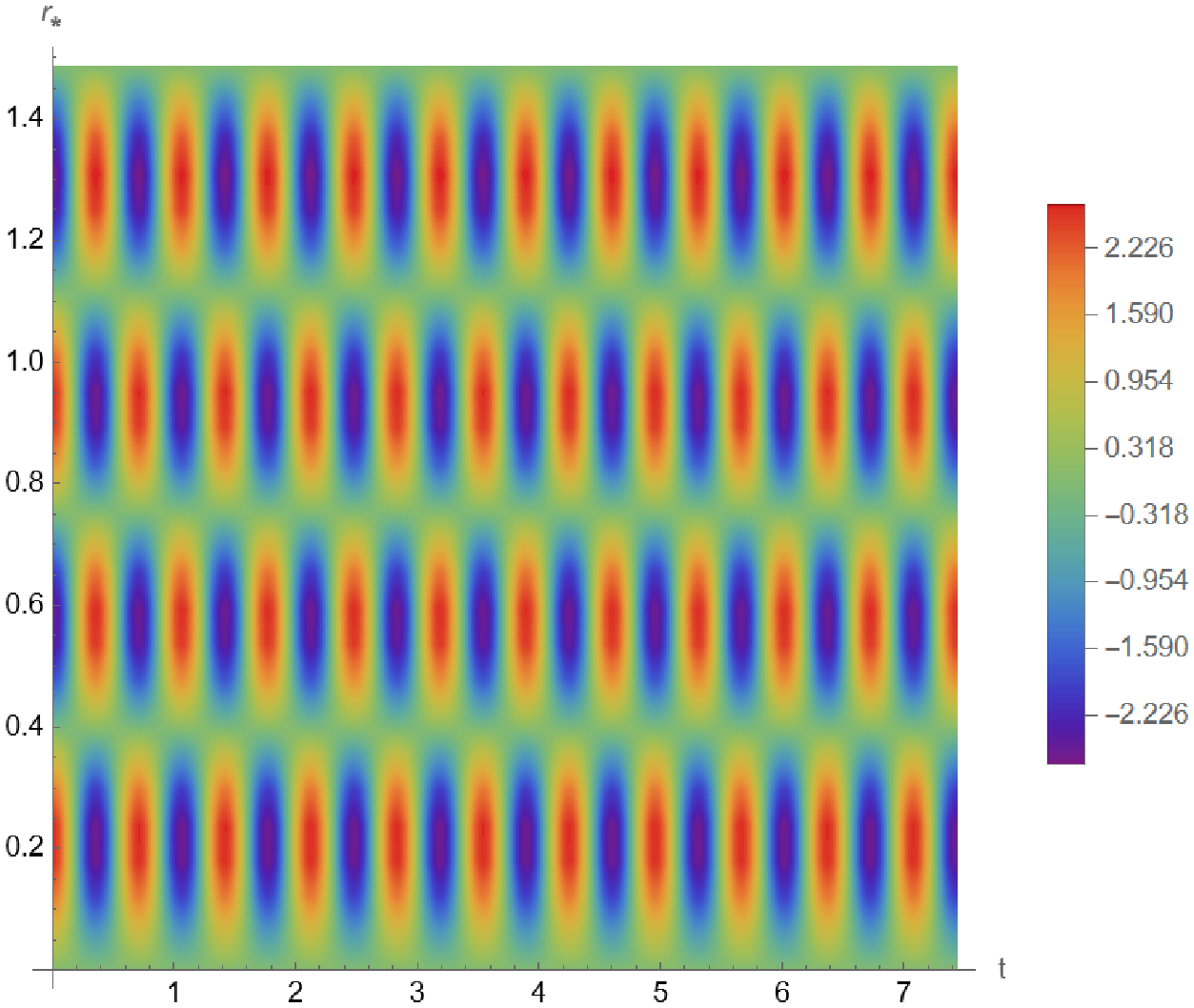}
\caption{Standing wave solutions of the axial perturbations in the RN-AdS metric with a singularity for overtones $n=0$ (upper-left), $n=1$ (upper-right), $n=2$ (bottom-left), and $n=3$ (bottom-right), where $M=3$, $Q=\sqrt{5}$, $\Lambda=-3$, and $L=3$.} \label{fig3}
\end{figure}

\begin{figure}
\includegraphics[width=8cm]{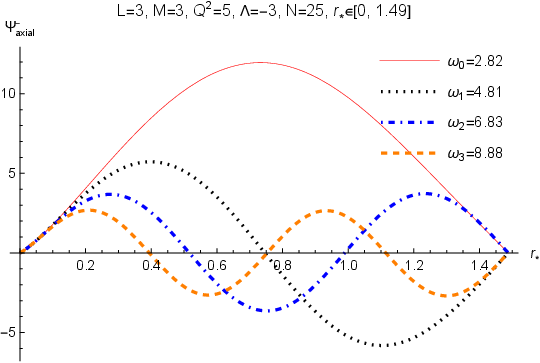}
\caption{Spatial profiles of the standing wave solutions of the axial perturbations with overtones $n=0, 1, 2$, and $3$.} \label{fig4}
\end{figure}

The resultant spatial-temporal evolutions of the initial perturbations are presented in Figs.~\ref{fig2}-\ref{fig4}. 
For illustration purposes, the radial coordinates in both scenarios are transformed into the interval $(0,1)$.
For the extreme RN-AdS black hole metric, it is defined as $x=1-1/r$, while for the metric with a naked singularity, we adopt $y=r_*/r_\text{*inf}$.

In Fig.~\ref{fig2}, we show the spatial-temporal evolutions of the axial perturbations of a Gaussian wave package in the extreme RN-AdS black hole metric and in that with a naked singularity.
The case for RN-AdS black hole is shown in the left panel, where one assumes $M=27/16$, $Q=\sqrt{11/8}$, $\Lambda=-3$, $L=3$, and thus the outer horizon $r_h=1$.
The initial perturbation is given by $\Psi(t=0)=e^{-100(r_*+1)^2}$ and $\partial_t\Psi(t=0)=0$, and we choose $\Delta x = 2\Delta t = 0.01$ in the finite difference method.
The case for naked singularity is given in the right panel, where one takes $M=3$, $|Q|=\sqrt{5}$, $\Lambda=-3$, and $L=3$.
The initial perturbation is given by $\Psi(t=0)=e^{-1250(r_*-\frac12)^2}$ and $\partial_t\Psi(t=0)=0$, and we adopt $\Delta r_* = 2\Delta t = \frac{r_*^\infty}{N_{r_*}}$ in the finite difference method. 
The total grid points in $r_*$ and $t$ coordinates are $N_{r_*}=10^3$ and $N_t=10^4$, respectively.
Distinct features are observed in the spatial-temporal evolutions of the two scenarios.
The waveform is primarily dissipative while non-oscillative for the black hole quasinormal modes shown in the left panel. 
This is because the waveform is constituted mainly by the underlying quasinormal modes that are purely imaginary.
A half-wave loss occurs when the waveform is reflected at $x=1$ as the amplitude flips its sign. 
This is a direct consequence of the asymptotical divergence of the effective potential at the spatial infinity.
On the other hand, the perturbations never reach the horizon at $x=0$ due to the infinite redshift.
The initial perturbations bounce off the spatial infinity only once for the above reason.
The right panel of Fig.~\ref{fig2} shows the scenario of a naked singularity, and it is observed that the waveforms do not diminish in time.
In the chosen spatial coordinate, the waveform travels essentially at constant velocity as it is repeatedly reflected between the boundaries, namely, the singularity $x=0$ and spatial infinity $x=1$. 
Again, the half-wave loss is observed at both boundaries, which can be attributed to the asymptotical properties of the effective potential shown in Fig.~\ref{fig1}.

The waveforms shown in the right panel of Fig.~\ref{fig2} can be decomposed as a summation of standing waves, shown in Fig.~\ref{fig3}.
In the latter, we present the profiles of the first few overtones of standing waves. 
The finite difference method is carried out using $\Delta r_*=2\Delta t=\frac{r_*^\infty}{N_{r_*}}$ and the total grid points in $r_*$ and $t$ axes are $N_{r_*}=10^3$ and $N_t=10^4$, respectively.
As standing waves, the perturbations oscillate in time for a given spatial point.
Meanwhile, a given phase, roughly a given color fragment, is localized as it only stretches along the spatial direction.
For these waveforms, it is observed that the boundaries always correspond to the nodes, while modes with different overtones can be identified by the number of antinodes.
    
We further explore the spatial waveform and the quasinormal frequencies pertaining to the standing wave solutions obtained in the section by employing the matrix and Prony methods.
The matrix method~\cite{agr-qnm-lq-matrix-01,agr-qnm-lq-matrix-02,agr-qnm-lq-matrix-03,agr-qnm-lq-matrix-04,agr-qnm-lq-matrix-05,agr-qnm-lq-matrix-06,agr-qnm-lq-matrix-08,agr-qnm-lq-matrix-11} is an approach for reformulating QNM problems into matrix equations for complex frequencies. 
The method, akin to the continued fraction method, mainly differs in the choice of grid points for the waveform expansion~\cite{agr-qnm-lq-matrix-01} and can be viewed as a further generalization of the method proposed in~\cite{agr-qnm-continued-fraction-09}.
It offers versatility, addressing spherically symmetric cases~\cite{agr-qnm-lq-matrix-02}, extending to metrics with axial symmetry~\cite{agr-qnm-lq-matrix-03}, and accommodating various boundary conditions~\cite{agr-qnm-lq-matrix-04}. 
Its competence extends to systems with coupled degrees of freedom~\cite{agr-qnm-lq-matrix-07} or coupled master equation sectors~\cite{agr-qnm-lq-matrix-03}. 
Furthermore, it has been adapted for dynamic black hole spacetimes~\cite{agr-qnm-lq-matrix-05} and recently generalized~\cite{agr-qnm-lq-matrix-06} for handling effective potentials with discontinuity.
The Prony method~\cite{agr-qnm-55} is a powerful tool in data analysis and signal processing. 
The method is implemented by turning a non-linear minimization problem into that of linear least squares in matrix form, which is particularly useful for extracting complex frequencies from a regularly spaced time series.
For the results presented below, the convergence of the numerical values has been confirmed by using different grid sizes.

\begin{table}[ht]
\caption{\label{TableI} The obtained frequencies $\omega$ using the matrix and Prony methods.
The calculations are carried out using the parameters $M=3$, $Q=\sqrt{5}$, $\Lambda=-3$, and $L=3$.
The matrix method is employed by using $N=25$ grid points.
The finite difference method uses the resolution $\Delta r_*=0.015$ and $\Delta t=0.0074$.
Subsequently, the Prony method is used to extract the frequencies by applying to the time interval $[0, 7.44]$.}
\begin{tabular}{c c c }
\hline\hline
n &~~~matrix method~~~&~~~Prony method~~~\\
\hline
$0$ & $2.82459+0.00000i $ & $2.82523 - 0.000182105 i$ \\
$1$ & $4.80679+0.00000i$ & $4.81311 - 0.0120591 i$ \\
$2$ & $6.82717+0.00000i$ & $6.82831 - 0.0000433076 i$ \\
$3$ & $8.88166+0.00000i$ & $8.88486 + 0.000215095 i$ \\
\hline\hline
\end{tabular}
\end{table}

In Fig.~\ref{fig4}, the corresponding spatial profiles of the first few overtones for the axial perturbations are depicted, and the corresponding frequencies are also indicated.
Because the system is not dissipative, the obtained frequencies are real numbers.
These results are manifestly consistent when compared against those obtained using the Prony method to extract the oscillation frequencies from the waveform obtained earlier.
The comparison is presented in Tab.~\ref{TableI}.
The frequencies obtained using two independent approaches reasonably agree with each other up to a discrepancy of less than 1\%.

\section{A comparison between the quasinormal oscillations and standing waves}\label{section4}
\renewcommand{\theequation}{5.\arabic{equation}} \setcounter{equation}{0}

In this section, we delve into the discussions about the difference between the quasinormal oscillations and standing wave solutions established in the previous section.
In particular, we compare two scenarios: an extreme RN-AdS black hole and a spacetime where a naked singularity is marginally formed.

The metric with a naked singularity falls back to that of a charged black hole in the asymptotically AdS spacetime when the condition Eq.~\eqref{condSing} is not satisfied.
For an RN-AdS black hole, the spacetime is known to be dissipative and stable~\cite{agr-qnm-47}, where the initial perturbations are expected to disperse and eventually be absorbed by the charged black hole.
In what follows, we consider the scenario where, on the one hand, the black hole is extreme; on the other hand, a naked singularity is marginally formed.
Indeed, for RN-AdS black hole spacetimes, while the metrics are stable against perturbations, strong cosmic censorship~\cite{agr-cosmic-censorship-01} might be broken for near extreme black holes~\cite{agr-cosmic-censorship-55, agr-cosmic-censorship-56, agr-cosmic-censorship-39, agr-cosmic-censorship-57, agr-cosmic-censorship-65}.
In this context, it is arguable whether one might turn an extreme black hole into a naked singularity by injecting a test particle into the black hole~\cite{agr-cosmic-censorship-05, agr-cosmic-censorship-15, agr-cosmic-censorship-16, book-general-relativity-Hartle}.
Therefore, under such a circumstance, the physical properties of the gravitational system, inclusively the quasinormal modes, have been an intriguing topic explored in the literature~\cite{agr-cosmic-censorship-13, agr-cosmic-censorship-15, agr-cosmic-censorship-16, agr-bh-horizon-15, agr-bh-horizon-16, agr-qnm-39, agr-qnm-47, agr-qnm-48, agr-qnm-49}.

For an extreme black hole, we choose $M=3$, $Q=2$, $\Lambda=-3$ and $L=3$, so that two horizons become degenerated and are given by
\begin{equation}\label{ExbhPar}
r_c=r_\pm = 1
\end{equation}
is a one-way membrane from the outside.
In the vicinity of such an extreme black hole, one may consider the marginal metrics with and without the horizon.
At first glimpse, the metric should be identical for $r>r_c$.
However, it turns out not to be the case and subsequently leads to a drastic change between the two fundamental modes, potentially reflecting the distinction in the physical natures of the two scenarios.

Specifically, for a black hole, the flux of the waveform must be ingoing at the horizon.
Nonetheless, as the black hole approaches an extreme one, it was numerically indicated that the obtained quasinormal frequency has a vanishing imaginary part~\cite{agr-qnm-47}, different from those for asymptotically flat spacetimes~\cite{agr-qnm-39, agr-qnm-48, agr-qnm-49}.
In the literature, it has been argued~\cite{agr-cosmic-censorship-55, agr-cosmic-censorship-57, agr-cosmic-censorship-65} that the magnitude of the imaginary part of the quasinormal frequencies implies that the corresponding relaxation rate of the collapsed charged fields is slow enough to lead to significant mass inflation.
The latter, in turn, destabilizes the dynamically formed inner Cauchy horizon and breaks the strong cosmic censorship~\cite{agr-cosmic-censorship-13, agr-cosmic-censorship-56}.
For the present case, the original one-way membrane located at the coordinate $r_c$ turns out to be a node of the standing waves, whose waveform is stable against axial metric perturbations.

\begin{figure*}
\includegraphics[width=8cm]{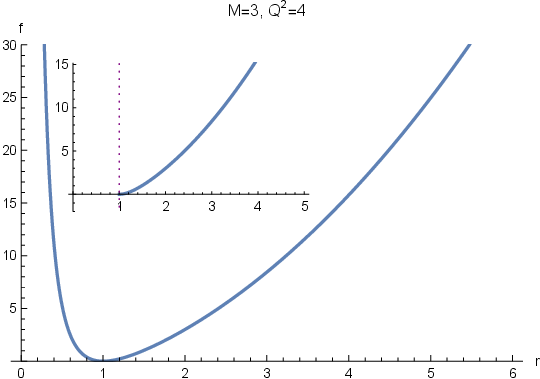}
\includegraphics[width=8cm]{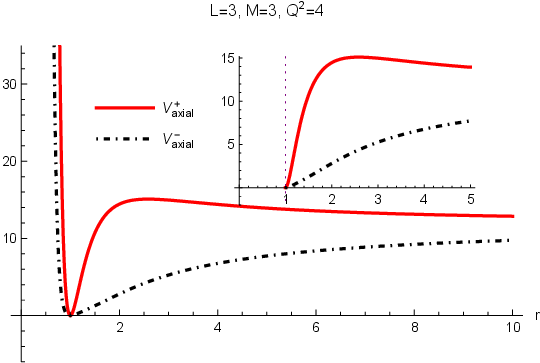}
\caption{The metric function $f(r)$ (left) and the effective potentials $V_\text{axial}$ (right) for the extreme RN-AdS spacetime.} \label{fig5}
\end{figure*}

\begin{figure}
\includegraphics[width=8cm]{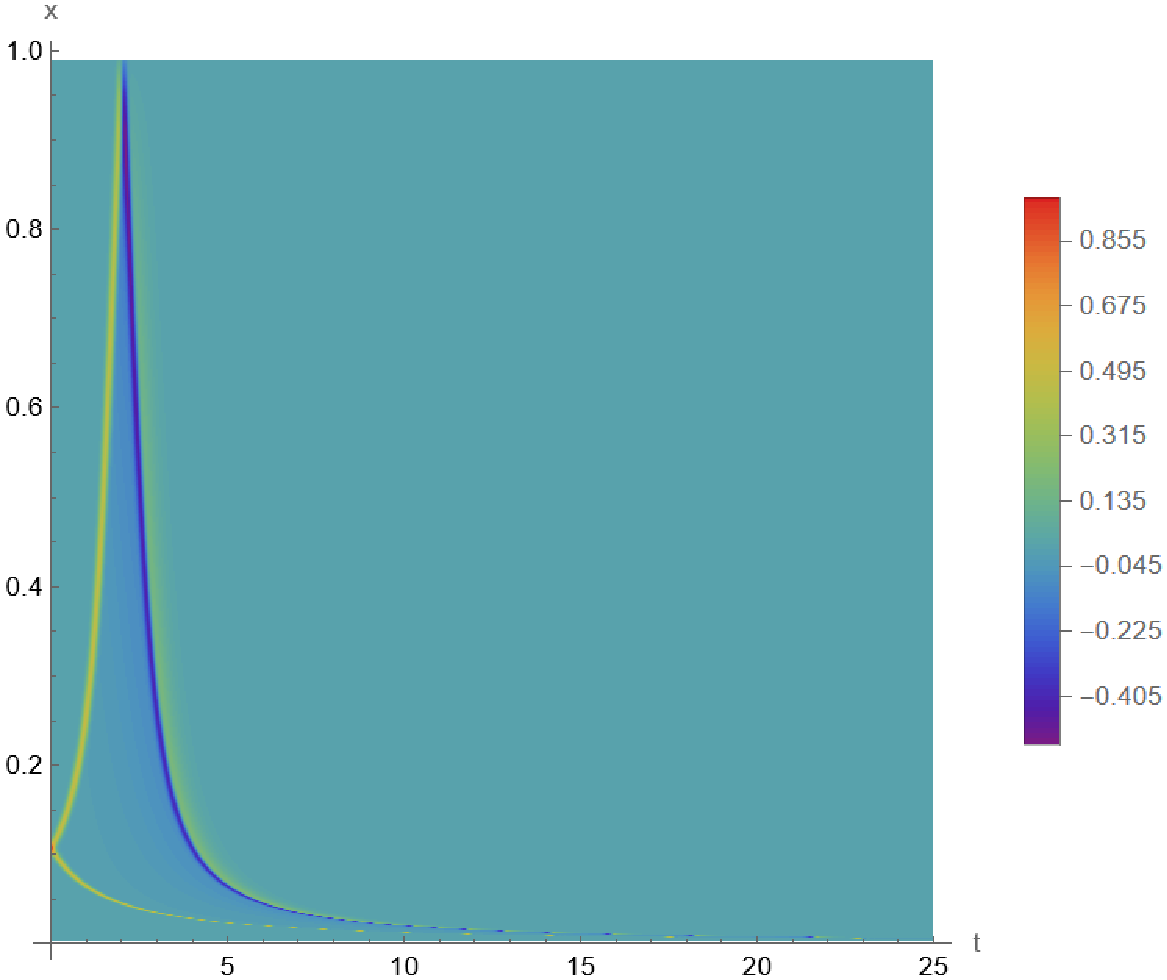}\includegraphics[width=8cm]{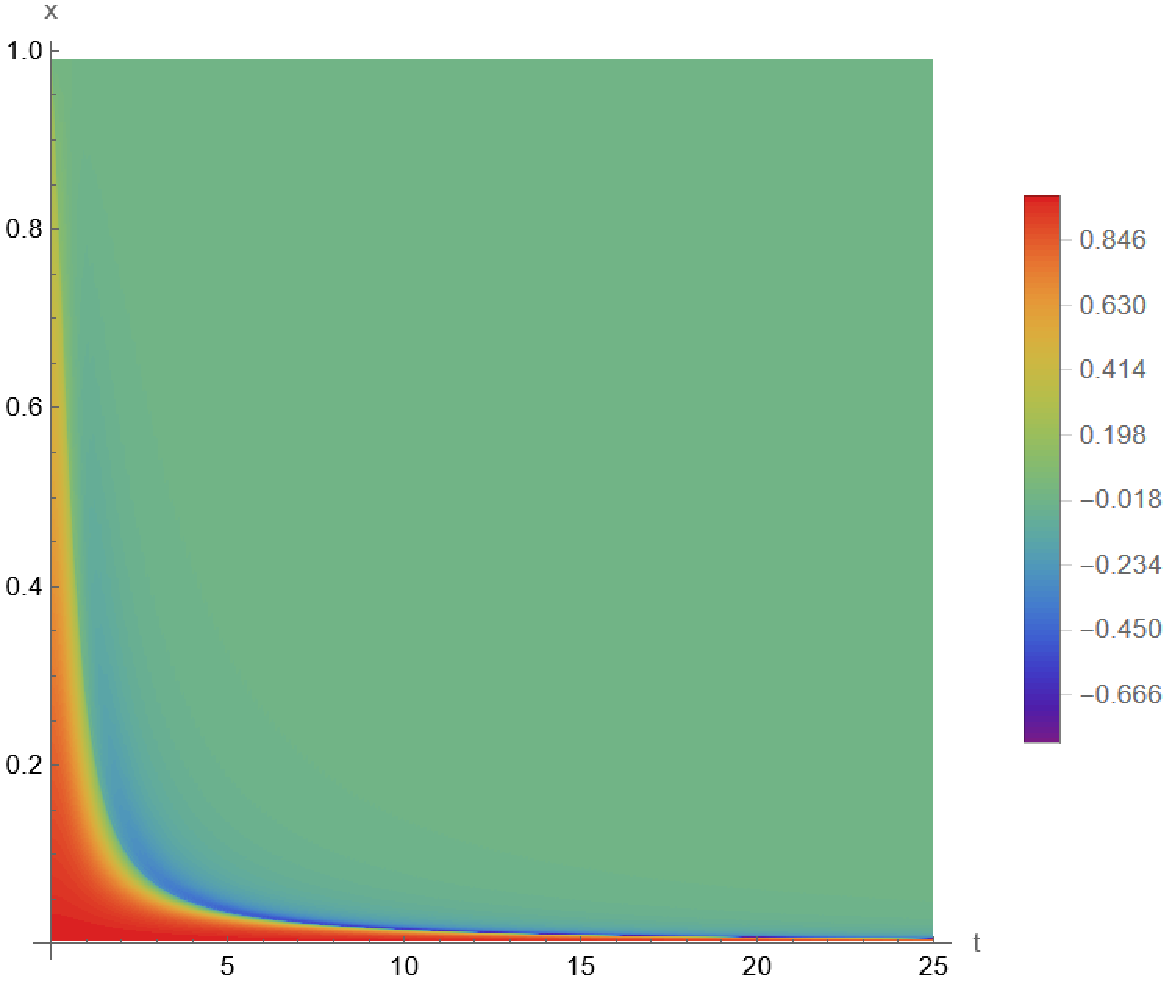}
\caption{The space-temporal evolution in the extreme RN-AdS black hole metric of the initial axial perturbations of a Gaussian form (left) and that of the fundamental quasinormal mode (right).
The calculations are carried out using the parameters $M=3$, $Q=2$, $\Lambda=-3$, and $L=3$.} \label{fig6}
\end{figure}

\begin{figure}
\includegraphics[width=8cm]{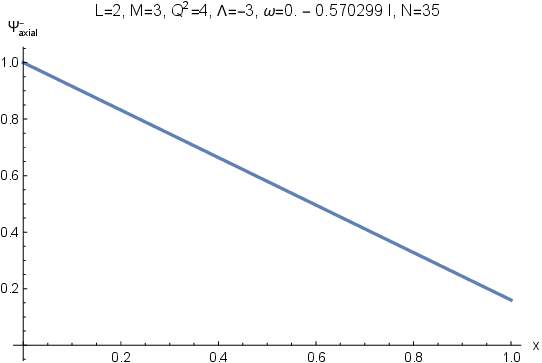}\includegraphics[width=8cm]{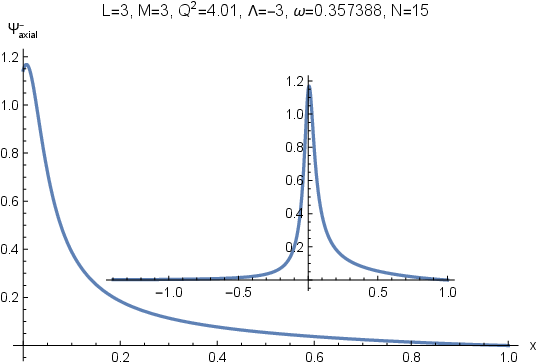}
\caption{The spatial profile of the fundamental modes in the extreme black hole metric shown in Fig.~\ref{fig6} (left) and that when a naked singularity marginally formed (right) shown in Fig.~\ref{fig8}.} \label{fig7}
\end{figure}

\begin{figure}
\includegraphics[width=8cm]{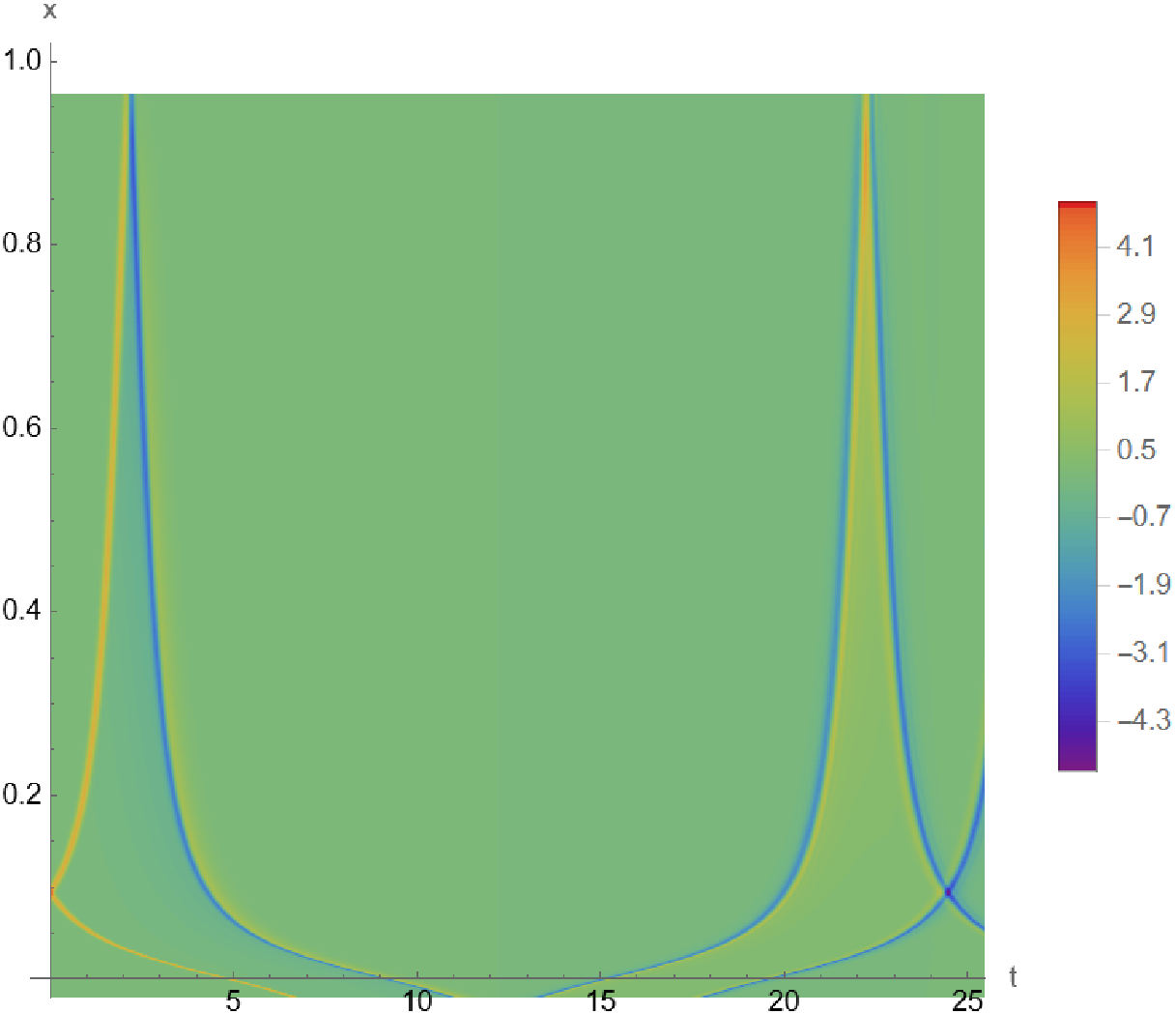}\includegraphics[width=8cm]{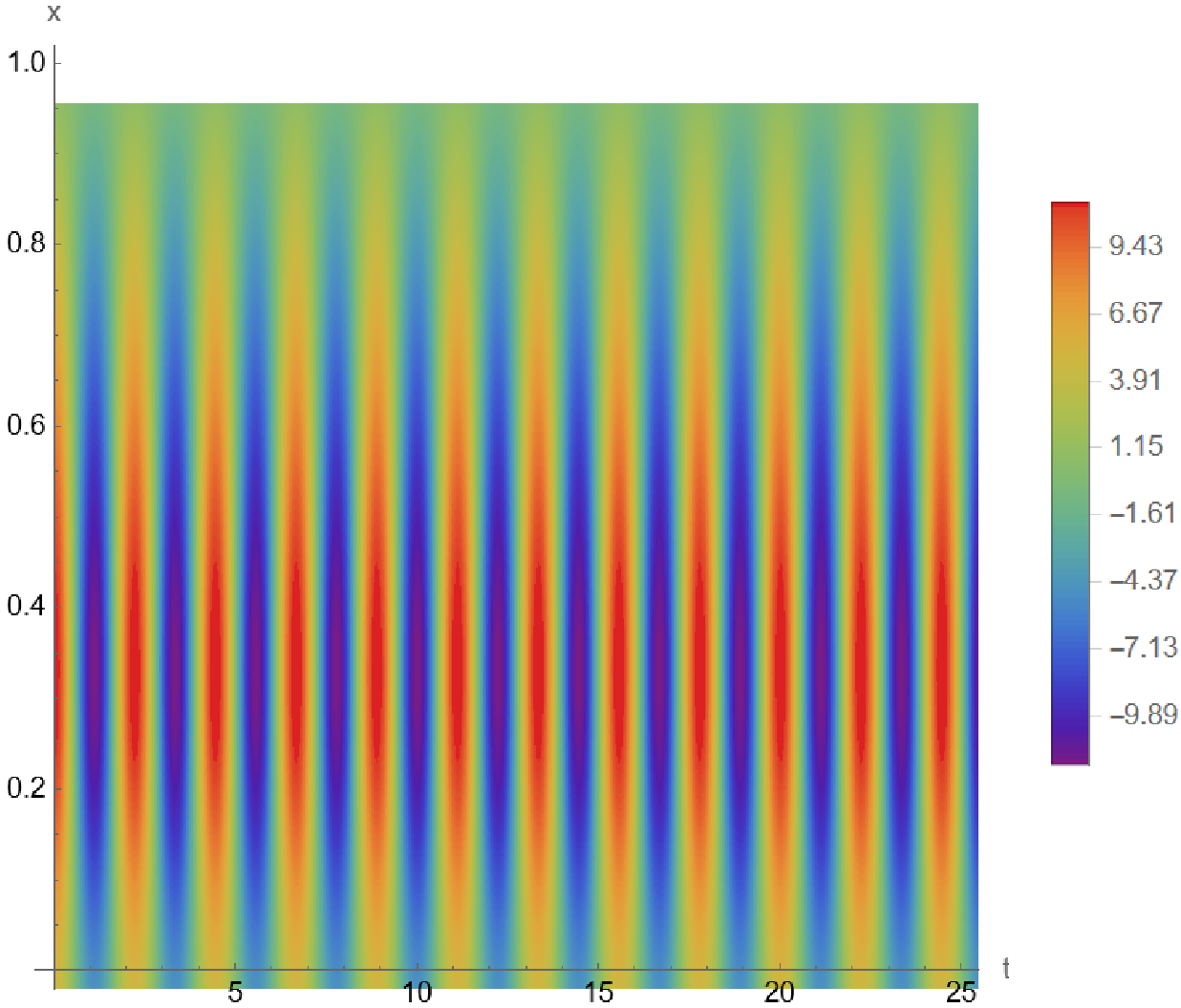}
\caption{The space-temporal evolution of axial perturbations in the RN-AdS spacetime with a naked singularity marginally formed and that of the fundamental mode of the standing wave.
The calculations are carried out using the parameters $M=3$, $Q=\sqrt{4.01}$, $\Lambda=-3$, and $L=3$. 
To compare against Fig.~\ref{fig6}, only the spatial region $r>r_c$ is presented, and the definition of the renormalized radial coordinate $x$ is identical in both figures.} \label{fig8}
\end{figure}

Our results are presented in Figs.~\ref{fig5}-\ref{fig8}.
The metric function and effective potential of extreme RN-AdS black hole metric are shown in Fig.~\ref{fig5}.
The corresponding space-temporal evolutions of an arbitrary initial Gaussian wave package and the fundamental mode of an extreme black hole are shown in Fig.~\ref{fig6}.
The initial perturbations in the extreme black hole are given by $\Psi(t=0)=e^{-100(r_*+2)^2}$ and $\Psi'(t=0)=0$, while the frequency of the fundamental modes is $\omega_\text{axial}=-0.570299i$.
For the finite difference method, we adopt $\Delta x=2\Delta t=0.01215$, and the total grid points $100$ and $3600$ in the spatial and temporal axes.
In the left panel of Fig.~\ref{fig6}, we see that the evolution of an arbitrary initial perturbation is qualitatively similar to that shown in the left panel of Fig.~\ref{fig2}, it is reflected at infinity while dissipates and absorbed by the horizon.
However, the fundamental mode shown on the right panel of Fig.~\ref{fig6} is suppressed in time without any temporal oscillations.
It possesses a non-trivial spatial distribution explicitly shown in the left panel of Fig.~\ref{fig7}, which remains unchanged in time.

As a comparison, in Fig.~\ref{fig8}, we show the spatial-temporal evolutions in the metric when a naked singularity is marginally formed.
The initial perturbations in the black hole are given by $\Psi(t=0)=6e^{-100(r_*-10)^2}$ and $\Psi'(t=0)=0$. 
The frequency for the fundamental mode is $\omega_\text{axial}=2.82$.
We adopt $\Delta r_*=2\Delta t=\frac{r_*^\infty}{100}$ for the finite difference method.
The total grid points in the spatial and temporal axes are $100$ and $3600$, respectively.
For the two cases, we note that the effective potential possesses the same limit at spatial infinity
\begin{equation}
\lim\limits_{r\to \infty}\lim\limits_{r_\pm\to r_c\pm 0}V(r) = \lim\limits_{r_\pm\to r_c\pm 0}\lim\limits_{r\to \infty}V(r) = L(L+1) .
\end{equation}
By comparing Figs.~\ref{fig6} and~\ref{fig8}, although the initial evolutions of a Gaussian waveform are similar, they are distinct.
For an extreme black hole, the system is dissipative.
The fundamental mode's waveform is finite at both boundaries with a finite flux.
When a naked singularity is marginally formed, the waveform is stationary, where the fundamental mode vanishes at the boundaries.

\section{Concluding remarks}\label{section5}
\renewcommand{\theequation}{5.\arabic{equation}} \setcounter{equation}{0}

To summarize, this work delves into the presence of standing wave solutions in the RN-AdS spacetime with a naked singularity. 
We identify these solutions as a unique subset of quasinormal modes bearing peculiar attributes. 
They are characterized by the frequencies with vanishing imaginary parts, contributing to their persistence over extensive timeframes. 
Additionally, regarding the spacetime waveform evolution, these modes primarily exhibit stationary behavior.
We corroborate the existence of such modes via numerical calculations using the finite difference method.
The Prony method is employed to extract the frequencies and compare the obtained results against those inferred through the matrix method.
These QNMs are shown to be drastically different from those of an extreme RN-AdS black hole. 
We believe our study offers some insights for the studies regarding metric perturbations.

\section*{acknowledgments}

National Natural Science Foundation of China (NNSFC) under contract No. 42230207 and the Fundamental Research Funds for the Central Universities, China University of Geosciences (Wuhan) with No. G1323523064.
We also gratefully acknowledge the financial support from
Funda\c{c}\~ao de Amparo \`a Pesquisa do Estado de S\~ao Paulo (FAPESP),
Funda\c{c}\~ao de Amparo \`a Pesquisa do Estado do Rio de Janeiro (FAPERJ),
Conselho Nacional de Desenvolvimento Cient\'{\i}fico e Tecnol\'ogico (CNPq),
Coordena\c{c}\~ao de Aperfei\c{c}oamento de Pessoal de N\'ivel Superior (CAPES),
A part of this work was developed under the project Institutos Nacionais de Ci\^{e}ncias e Tecnologia - F\'isica Nuclear e Aplica\c{c}\~{o}es (INCT/FNA) Proc. No. 464898/2014-5.
This research is also supported by the Center for Scientific Computing (NCC/GridUNESP) of S\~ao Paulo State University (UNESP).

\bibliographystyle{h-physrev}
\bibliography{references_qian, references_add}

\end{document}